%% file: CIKM2021.tex
\PassOptionsToPackage{table,xcdraw}{xcolor}

\documentclass[sigconf,authorversion]{acmart}

\usepackage[table,xcdraw]{xcolor}
\usepackage{multicol}
\usepackage{multirow}
\usepackage[normalem]{ulem}
\useunder{\uline}{\ul}{}
\usepackage{hyperref}
\usepackage{graphicx}
\usepackage{subfigure}
\usepackage{balance}

\copyrightyear{2021}
\acmYear{2021}
\setcopyright{acmcopyright}\acmConference[CIKM '21]{Proceedings of the 30th ACM International Conference on Information and Knowledge Management}{November 1--5, 2021}{Virtual Event, QLD, Australia}
\acmBooktitle{Proceedings of the 30th ACM International Conference on Information and Knowledge Management (CIKM '21), November 1--5, 2021, Virtual Event, QLD, Australia}
\acmPrice{15.00}
\acmDOI{10.1145/3459637.3482297}
\acmISBN{978-1-4503-8446-9/21/11}

\begin{document}
\fancyhead{}
\title{SimpleX: A Simple and Strong Baseline for Collaborative Filtering}

\author{Kelong Mao$^{1\ast}$,$\:$ Jieming Zhu$^{2\ast}$,$\:$  Jinpeng Wang$^3$,$\:$  Quanyu Dai$^2$,$\:$  Zhenhua Dong$^2$}
\author{Xi Xiao$^{3\star}$,$\:$  Xiuqiang He$^2$}
%\thanks{ Part of the work was done during the internship at Huawei Noah's Ark Lab when the author studied at Tsinghua University.}
\thanks{$^\ast$ Both authors contributed equally to the work.}
\thanks{$^\star$ Corresponding Author}

\affiliation{\vspace{2ex}$^1$Gaoling School of AI, Renmin University of China}
\affiliation{$^2$Huawei Noah's Ark Lab, Shenzhen, China}
\affiliation{$^3$Tsinghua Shenzhen International Graduate School, Tsinghua University, China}

\affiliation{\vspace{1ex}\texttt{mkl@ruc.edu.cn, jiemingzhu@ieee.org, wjp20@mails.tsinghua.edu.cn, daiquanyu@huawei.com \\
dongzhenhua@huawei.com,  xiaox@sz.tsinghua.edu.cn, hexiuqiang1@huawei.com}\vspace{2ex}}

\renewcommand{\authors}{Kelong Mao, Jieming Zhu, Jinpeng Wang, Quanyu Dai, Zhenhua Dong, Xi Xiao, Xiuqiang He}
% \author{Jieming Zhu}
% \authornotemark[2]
% \affiliation{Huawei Noah's Ark Lab}
% \email{jmzhu@ieee.org}

% \author{Jinpeng Wang}
% \affiliation{Tsinghua University}
% \email{wjp20@mails.tsinghua.edu.cn}

% \author{Quanyu Dai}
% \affiliation{Huawei Noah's Ark Lab}
% \email{daiquanyu@huawei.com}

% \author{Zhenhua Dong}
% \affiliation{Huawei Noah's Ark Lab}
% \email{dongzhenhua@huawei.com}

% \author{Xi Xiao}
% \affiliation{Tsinghua Shenzhen International Graduate School, Tsinghua University}
% \email{xiaox@sz.tsinghua.edu.cn}

% \author{Xiuqiang He}
% \affiliation{Huawei Noah's Ark Lab}
% \email{hexiuqiang1@huawei.com}

% https://www.overleaf.com/8426897472wkzxmswmkjsp

\begin{abstract}
Collaborative filtering (CF) is a widely studied research topic in recommender systems. The learning of a CF model generally depends on three major components, namely interaction encoder, loss function, and negative sampling. While many existing studies focus on the design of more powerful interaction encoders, the impacts of loss functions and negative sampling ratios have not yet been well explored. In this work, we show that the choice of loss function as well as negative sampling ratio is equivalently important. More specifically, we propose the {cosine contrastive loss} (CCL) and further incorporate it to a simple unified CF model, dubbed SimpleX. Extensive experiments have been conducted on 11 benchmark datasets and compared with 29 existing CF models in total. Surprisingly, the results show that, under our CCL loss and a large negative sampling ratio, SimpleX can surpass most sophisticated state-of-the-art models by a large margin (e.g., max 48.5\% improvement in NDCG@20 over LightGCN). We believe that SimpleX could not only serve as a simple strong baseline to foster future research on CF, but also shed light on the potential research direction towards improving loss function and negative sampling. Our source code will be available at \textcolor{magenta}{\url{https://reczoo.github.io/SimpleX}}.
%Collaborative filtering (CF) is a widely studied research topic in recommender systems. The effectiveness of a CF model depends on three major components, namely interaction encoder, loss function and the negative sampling strategy. While many research efforts have been made on the first two components i.e., designing better encoder structures and enhancing negative samples, few studies focus on the choice of loss functions and its collocation with negative sampling. In this work, we reveal that such a plausible setting is usually sub-optimal, and suitable collocation of loss functions and negative sampling strategies can largely boost the performance of existing CF models. In particular, we propose the use of \textit{cosine contrastive loss} together with more negative samples. In some cases, even very simple CF models such as MF can beat the state-of-the-art models. Furthermore, we present three simple yet effective baseline CF models. We hope that our work could not only serve as strong baselines for future research on collaborative filtering, but also attract more research efforts toward better loss function design.
\end{abstract}

%It is common that only positive feedbacks (e.g., clicks) are available in practice, thus we focus on implicit collaborative filtering in this work. 


\begin{CCSXML}
<ccs2012>
<concept>
<concept_id>10002951.10003317.10003347.10003350</concept_id>
<concept_desc>Information systems~Recommender systems</concept_desc>
<concept_significance>500</concept_significance>
</concept>
<concept>
<concept_id>10002951.10003227.10003351.10003269</concept_id>
<concept_desc>Information systems~Collaborative filtering</concept_desc>
<concept_significance>500</concept_significance>
</concept>
</ccs2012>
\end{CCSXML}

\ccsdesc[500]{Information systems~Recommender systems}
\ccsdesc[500]{Information systems~Collaborative filtering}

\keywords{Recommender systems; collaborative filtering; graph neural networks; contrastive loss}

\maketitle

\input{sections/1_introduction.tex}
\input{sections/2_background.tex}

\input{sections/3_approach.tex}
\input{sections/4_experiment.tex}
\input{sections/5_relatedwork.tex}
\input{sections/6_conclusion.tex}

\bibliographystyle{ACM-Reference-Format}
\balance
\bibliography{CIKM2021}
\end{document}

%% file: sections/1_introduction.tex
\section{Introduction}
\label{sec::intro}

Nowadays, personalized recommendation is ubiquitous in various applications, such as video recommendation in YouTube~\cite{YouTubeNet}, product recommendation in Amazon~\cite{amazonrecsys}, and news recommendation in Bing~\cite{NPA}. The goal of recommendation is to predict whether a user will interact (e.g., click or purchase) with an item and thus help users discover potential items of interests. Collaborative filtering (CF)~\cite{CF_survey} is a fundamental task in recommendation that leverages the collaborative information among users and items to predict users' preferences on candidate items. The simplicity and effectiveness make it one of the most popular techniques in recommender systems.

%Over the past two decades, remarkable progress has been made towards the CF, ranging from heuristic-based models~\cite{sarwar2001item, lops2011content}, to latent factor models based on low-rank matrix factorization (MF)~\cite{MF, koren2008factorization}, and to more recent deep learning-based models. 

Generally, the learning process of a CF model can be separated to three major components, including interaction encoder, loss function, and the negative sampling strategy used when only positive (i.e.,implicit) feedbacks are available. Most existing studies focus on the design of more powerful interaction encoders to capture collaborative signals among users and items. Especially, the prevalence of deep learning motivates a rich line of work that applies various neural networks to CF, including multi-layer perceptrons (MLPs)~\cite{NeuMF,YouTubeNet}, auto-encoders~\cite{MVAE}, attention networks~\cite{ACF}, transformers~\cite{transformerRec}, graph neural networks (GNNs)~\cite{LightGCN}, and so on. Nevertheless, these models tend to become more and more complex to show performance improvements. This somehow limits their practical applicability in industrial recommender systems that demand high efficiency. %As far as we know, simple models, such as MF~\cite{?} and its deep variant YouTubeNet~\cite{?}, are still the most popular industrial solutions due to their effectiveness and simplicity. 
% convolutional neural networks (CNNs)~\cite{ConvMF}, recurrent neural networks (RNNs)~\cite{GRU4Rec},

On the contrary, few research efforts have been devoted to investigating the impacts of the latter two components. Specifically, while multiple loss functions have been used in CF, such as Bayesian personalized ranking (BPR) loss~\cite{BPR}, binary cross-entropy loss~\cite{NeuMF}, softmax cross-entropy loss~\cite{YouTubeNet}, pairwise hinge loss~\cite{CML}, and mean square error loss~\cite{ENMF},
% early work~\cite{BPR} proposes the Bayesian personalized ranking (BPR) loss, NeuMF~\cite{NeuMF} takes the binary cross-entropy loss, YouTubeNet~\cite{YouTubeNet} employs the softmax cross-entropy loss for multi-label classification, 
% and CML~\cite{CML} applies the pairwise hinge loss in metric learning. 
there is still a lack of systematic evaluation and comparisons among different loss functions. Furthermore, many recent GNN-based studies~\cite{NGCF,LightGCN,DGCF,NIA-GCN,BGCF} experiment with the BPR loss~\cite{BPR} and simply set the negative sampling ratio to a small value (i.e., sampling 1 or 10 negative samples per positive user-item pair). In this way, they can justify the superiority of their proposed interaction encoders, but they neglect the importance of loss functions and negative sampling in the learning of CF models. 

In fact, we empirically observed that training with the BPR loss and a small negative sampling ratio results in inferior results for many CF models. In this paper, we show that choosing a suitable loss function and a proper number of negative samples plays an equal or more important role than an interaction encoder. Towards this goal, we systematically compare multiple commonly-used loss functions and also investigate the impact of negative sampling ratio on each loss function. Moreover, inspired by the widely used contrastive loss~\cite{ContrastiveLoss1,ContrastiveLoss2} in computer vision, we propose a cosine contrastive loss (CCL) tailored for CF. Our CCL loss optimizes the embedding by maximizing the cosine similarity of a positive user-item pair, while minimizing the similarity of a negative pair to a certain margin. Surprisingly, we found that even a simple model (e.g., MF), if paired with our proposed CCL loss, is sufficient to surpass many sophisticated state-of-the-art models.
% Meanwhile, many interaction encoders tend to achieve a similar-level performance under the CCL loss.

%, which largely reduces the performance gap brought by a well-designed interaction encoder.

%different loss functions and sampling strategies lead to drastically different solutions for the same loss function.

These findings raise questions about whether the current baselines are strong enough to verify the performance improvements of the state-of-the-art CF models, and how much these sophisticated models have really improved. Our work aims to answer these questions. We argue that the current baselines might not be strong enough, which could mislead us to overestimate the real improvements of many new CF models. Instead of criticizing the contributions of any existing work, the main goal of our work is to build a simple and strong baseline model to foster future research on CF. 

In the design of SimpleX, we keep simplicity in mind and borrow ideas from several existing studies (e.g., average pooling in YouTubeNet~\cite{YouTubeNet}, attention in ACF~\cite{ACF}). We build Simplex as a unified model that integrates matrix factorization and user behaviour modeling. Specifically, it comprises a behavior aggregation layer (e.g., average pooling) to obtain a user's preference vector from the historically interacted items, and then fuses with the user embedding vector via a weighted sum. More importantly, SimpleX is optimized with our CCL loss and a large negative sampling ratio. Although the interaction encoder of SimpleX seems quite simple and might not be novel at all, we show that it could serve as a super-strong baseline model and have great potential for industrial applications because of its high efficiency.

For evaluation, we conduct comprehensive experiments on 11 benchmark datasets in total and compare with a total of 29 popular CF models of different types. The results show that SimpleX outperforms most sophisticated state-of-the-art methods by a large margin (up to 48.5\% improvement in NDCG@20 over LightGCN~\cite{LightGCN} on Amazon-Books). We also empirically compare the performance of six representative loss functions and investigate the impact of different negative sampling ratios on each loss function, which demonstrates the superiority of our proposed CCL loss for CF tasks. Furthermore, we evaluate the efficiency of SimpleX, which shows more than 10x speedup over the simplified GNN-based CF model, LightGCN~\cite{LightGCN}. We hope that our work could not only serve as a simple and strong baseline to foster future research on CF, but also attract more research efforts towards the co-design of interaction encoders, loss functions, and negative sampling strategies.

%Results validate that our MF-CCL and SimpleX can consistently achieve SOTA performance on various datasets and settings with very simple designs, which are robust to serve as strong baseline models for future CF research. 

%In contrast to the BPR loss, CCL can easily help improve model performance by increasing the negative sampling ratio Furthermore, which is hard to be achieved by BPR loss. 

The main contributions of our work are summarized as follows:
\begin{itemize}
    \item We highlight the importances of loss functions and negative sampling in CF, and propose the cosine contrastive loss accordingly.
   
    \item We present a simple and strong baseline model, SimpleX, which could even attain much better performance than most sophisticated state-of-the-art models.
    
    \item We perform experiments on 11 benchmark datasets and compare SimpleX with 29 existing CF models to show its superiority in terms of both effectiveness and efficiency. 
\end{itemize}

%% file: sections/2_background.tex
\section{Background and Related Work}\label{sec:cf}

%In this section, we first give a formulation of collaborative filtering algorithms, pointing out three important aspects in CF models' learning in detail, and then introduce four categories of CF models.
In this section, we first give a formulation of collaborative filtering  and point out three important aspects in CF modeling. We then summarize different categories of CF models.

\subsection{Formulation of CF}
% Collaborative filtering is one of the most widely used techniques to build intelligent recommender systems.
The research of collaborative filtering includes implicit CF and explicit CF. Implicit CF models learn from implicit feedback data, e.g., click, visit, and purchase, while explicit CF models learn from explicit feedbacks such as ratings. 
% In this work, we focus on implicit CF since it is more universal as the data in the real recommendation scenarios is usually implicit feedback. 
In this work, we focus on implicit CF since it is more common in real recommendation scenarios. 
Besides, it is also easy to transform explicit feedback to implicit feedback via binarization. 
In implicit CF, a matrix $\mathbf{Y}$ is used to denote the user-item interactions, where $y_{ui}=1$ if user \textit{u} has observed interaction with item \textit{i} and $y_{ui}=0$ otherwise.

As mentioned in Section~\ref{sec::intro}, 
we highlight three vital aspects that have a large impact to the learning process of CF models:

% \begin{itemize}
    % \item 
    % \noindent
    \textbf{(1) Interaction Encoder.} 
    % The function of the interaction encoder is to learn embeddings (or latent features) for each user and item, which capture collaborative signals that reflect the behavioral similarity between users (or items), from the interaction matrix. 
    The function of the interaction encoder is to learn embeddings for each user and each item, which capture collaborative signals in the interaction matrix that reflect the behavioral similarity between users (or items). 
    It is undoubtedly the core of CF models and has been well studied. We give a brief summary of interaction encoders in section~\ref{sec::sotacf}.
    
    % \item 
    % \noindent
    \textbf{(2) Loss Function.} 
    % Two types of loss functions are commonly used - pointwise loss and pairwise loss. 
    In general, there are two common types of loss functions in CF.
    Pointwise loss functions such as binary cross-entropy (BCE) and mean square error (MSE) treat the learning process as a binary classification or a regression task.
    Pairwise loss such as Bayesian personalized ranking loss (BPR) is optimized to make the similarities of positive user-item pairs larger than the negative ones.
    
    % \item 
    % \noindent
    \textbf{(3) Negative Sampling.} 
    Since there are a lot of unobserved entries, in most cases we need to perform negative sampling to improve training efficiency. A few studies have been made to improve the uniform random sampling for recommendation, including mining informative negative samples (e.g., RNS~\cite{reinforced_sampling}, and NBPO~\cite{NBPO}), tackling the selection bias of implicit user feedback (e.g., MSN~\cite{mixed_sampling}) and so on. 
    In this work, we mainly investigate the influence of the negative sampling ratio.
    The existing studies are complementary to our work and potential to be applied to our SimpleX model for further improvement.
% \end{itemize}

\subsection{Summary of representative CF methods}
\label{sec::sotacf}
% Extensive research on CF has been done over the past two decades. We summarize the recent representative CF methods into the following four categories:
We summarize representative CF methods into four categories:

\textbf{(1) MF-based methods.} 
% MF-based algorithms work by decomposing the user-item interaction matrix into the product of two lower dimensionality rectangular matrices which are user latent vectors and item latent vectors respectively, and use them for the prediction of unknown entries. 
Matrix factorization (MF) based algorithms decompose the user-item interaction matrix into two low-dimensional latent matrices for user and item representation. 
Due to its effectiveness, MF has been wildly studied in CF. 
% Due to its effectiveness, there have emerged many good MF-based methods. 
% For example, 
% Manotumruksa et al. propose GRMF~\cite{GRMF} that smooths MF through adding the graph Laplacian regularizer to introduce graph information. 
% Yang et al. devise a unified and efficient method called HOP-Rec~\cite{HOP-Rec} that incorporates both MF and graph-based models for implicit CF.
% Chen et al. design ENMF~\cite{ENMF} which is an efficient MF-based CF framework that relies on MSE loss function without the need of negative sampling for implicit CF.
Manotumruksa et al. proposed GRMF~\cite{GRMF} that smoothed MF through adding the graph Laplacian regularizer to introduce graph information. 
Yang et al. devised a unified and efficient method called HOP-Rec~\cite{HOP-Rec} that incorporated both MF and graph-based models for implicit CF.
Chen et al. designed ENMF~\cite{ENMF}, which is an efficient MF-based CF model with modified MSE loss function. It can be optimized efficiently without negative sampling for implicit feedback.
% It is worth mentioning that the thinking of MF, i.e., learning the latent vectors (a.k.a. embeddings) for users and items has become a normative paradigm now. 
% Vanilla MF only uses the latent vectors of users and items to fit the interaction matrix. 
    
\textbf{(2) Autoencoder-based methods.}
% Autoencoder-based CF methods leverages the autoencoder, which is a powerful generative model, to learn the embeddings of users and items. Such generative models can perform inductive recommendation (i.e., learning from one group of users while performing recommendation for other group users with the same candidate items) and help to solve the cold start problems. For example, 
% Liang et al. propose Mult-VAE~\cite{MVAE}, which applies variational autoencoder (VAE) for CF. 
% Ma et al. propose MacridVAE~\cite{MacridVAE} by disentangling user intents behind user-item and leveraging $\beta$-VAE to simulate the generative process of a user’s personal history interactions.
% Steck et al. design a linear model called EASE\textsuperscript{R}~\cite{EASE} that is geared toward sparse data in particular implicit feedback data for the recommendation.
Autoencoder-based CF methods leverage the autoencoder network architectures to learn item embeddings. Such models are suitable to perform inductive recommendation, i.e., learning from one group of users while performing recommendation for another group of users with the same candidate items. For example, 
Liang et al. proposed Mult-VAE~\cite{MVAE}, which applied variational autoencoder (VAE) for CF. 
Ma et al. proposed MacridVAE~\cite{MacridVAE} by disentangling user intents behind user-item and leveraging $\beta$-VAE to simulate the generative process of a user’s personal history interactions.
Steck et al. designed a linear model called EASE\textsuperscript{R}~\cite{EASE} that is geared toward sparse data, in particular implicit feedback data, for the recommendation.

\textbf{(3) GNN-based methods.}
Since the interaction data can be naturally modelled as a user-item bipartite graph, 
recent studies propose graph neural network (GNN) based CF models and report state-of-the-art performance.
% more and more GNN-based CF models are emerging 
% and have achieved much better performance than previous methods. 
% They modeled the recommendation as a link prediction problem between user nodes and item nodes, which can effectively capture higher-order collaborative signals through multi-layers message passing in GNNs. 
GNN-based methods model the recommendation as the link prediction task between user nodes and item nodes, 
% which can effectively capture higher-order collaborative signals through multi-layers message passing in GNNs. 
where the higher-order collaborative signals can be effectively captured through multi-layers message passing.
% We briefly introduce a part of representative GNN-based CF models. 
% Berg et al. propose GC-MC~\cite{GC-MC} by directly appling a one-layer GNN to the user-item bipartite graph with the weight-sharing technique to extract user and item latent features. 
Ying et al. proposed PinSage~\cite{PinSage} that improved GraphSage~\cite{hamilton2017inductive} to model the item-item relationships for Pinterest.
Wang et al. devised NGCF~\cite{NGCF} that explicitly encoded the collaborative signals as high-order connectivities by performing embedding propagation.
% Then He et al. find that NGCF's design is burdensome for CF, and propose LightGCN~\cite{LightGCN} by removing the feature transformation and non-linear activation in NGCF. 
He et al. proposed LightGCN~\cite{LightGCN}, which removed the feature transformation and non-linear activation in NGCF and improved both performance and efficiency.
These successful applications of GNN in recommendation further inspire many good studies, including BGCF~\cite{BGCF} which models the uncertainty in the user-item graph with bayesian graph neural networks, DGCF~\cite{DGCF} which models a distribution over intents for each user-item interaction, NIA-GCN~\cite{NIA-GCN} and NGAT4Rec~\cite{NGAT4Rec} that learn neighborhood relationships, and SGL-ED~\cite{SGL-ED}, DHCF~\cite{DHCF}, LCFN~\cite{LCFN}, and so on.

\textbf{(4) Others.}
We put methods that do not fall into the first three categories into this ``Others'' category. Here we list some representative models such as SLIM~\cite{SLIM} which is a simple linear model that combines the advantages of neighborhood- and model-based CF approaches, MLPs-based NeuMF~\cite{NeuMF} and YouTubet~\cite{YouTubeNet},  memory network-based CMN~\cite{CMN}, metric learning-based CML~\cite{CML}, and NBPO~\cite{NBPO} that leverages noisy-label robust learning techniques.

%% file: sections/3_approach.tex
\section{SimpleX}\label{sec:approach}
In this section, we first present our cosine contrastive loss and the SimpleX model architecture for CF. We then analyze its connections to other existing models.

\subsection{Cosine Contrastive Loss}
\label{approach:MF-CCL}

In the CF literature, many different loss functions have been employed, including BPR loss~\cite{BPR}, binary cross-entropy~\cite{NeuMF}, softmax cross-entropy~\cite{YouTubeNet}, pairwise hinge loss~\cite{CML}, etc. However, there is still a lack of a systematic comparison among them, leaving their effects on model performance not well understood. In this work, we not only make such a comparison, but also propose a new loss function for CF, namely cosine contrastive loss (CCL). Given a positive user-item pair ($u$, $i$) and a set of randomly sampled negative samples (i.e., $\mathcal{N}$), the CCL loss is expressed as follows:
\begin{equation}\label{equ:ccl}
    \mathcal{L}_{CCL}(u,i) = (1 - \hat{y}_{ui})+ \frac{w}{|\mathcal{N}|} \sum_{j \in \mathcal{N}}\operatorname{max}(0, \: \hat{y}_{uj} - m)
\end{equation}
where $\hat{y}_{ui}$ calculates the cosine similarity between the representation vectors of user $u$ and item $i$. $|\mathcal{N}|$ denotes the number of negative samples. $m$ is the margin to filter negative samples, which is usually set to 0$\sim$1. Intuitively, CCL is optimized to maximize the similarity between positive pairs and minimize the similarity of negative pairs below the margin $m$. $w$ is a hyper-parameter to control the relative weights of positive-sample loss and negative-sample loss. 

\textbf{Design Choices}. 
The formulation of CCL is simple and largely inspired by the widely used contrastive loss~\cite{ContrastiveLoss1,ContrastiveLoss2} in the computer vision tasks, such as face recognition and image retrieval. But we make several design choices that differ from most widely-used loss functions in CF and greatly facilitate model training. First, instead of applying dot product (e.g., in LightGCN~\cite{LightGCN}) or Euclidean distance (e.g., in CML~\cite{CML}) to measure the similarity (or distance) between a user-item pair, we choose to compute the cosine similarity between them. By applying L2 normalization on both representation vectors, cosine similarity only calculates the angle difference and thus 
avoid the effect of representation magnitude. This is favorable since the magnitude of a user/item representation could be strongly biased by its popularity in CF tasks. This is also similar to the calculation of word similarity in Word2Vec~\cite{word2vec}, where cosine similarity is usually used.

Second, when the number of negative samples becomes large, there usually exist many redundant yet uninformative samples. But existing loss functions (e.g., BPR~\cite{BPR}) treat every negative sample equivalently. As such, model training could be overwhelmed by these uninformative samples, which significantly degrade the model performance and also slows the convergence. In contrast, CCL alleviates this problem by using a proper margin $m$ to filter uninformative negative samples. Intuitively, uninformative negative samples will get zero loss in CCL when they have a small cosine similarity below the margin $m$. As a result, it helps automatically identify those hard negative samples with cosine similarity larger than $m$, and thus facilitates better training of the model. 

Third, we found that directly summing or averaging the loss terms of all negative samples could degrade the model performance, especially when the number of negative samples is large. This is partially due to the high imbalance between positive and negative samples (e.g., 1:1000 when $|\mathcal N|=1000$). We thus introduce a data-dependent weight $w$ to control the balance between positive loss and negative loss. We emphasize that it also achieves a similar effect to the confidence weight imposed on negative samples in weighted matrix factorization~\cite{WMF}.

\begin{figure}[!t]
    \begin{center}
    %\framebox[4.0in]{$\;$}
    % \fbox{\rule[-.5cm]{0cm}{4cm} \rule[-.5cm]{4cm}{0cm}}
    \includegraphics[width = 0.8\linewidth]{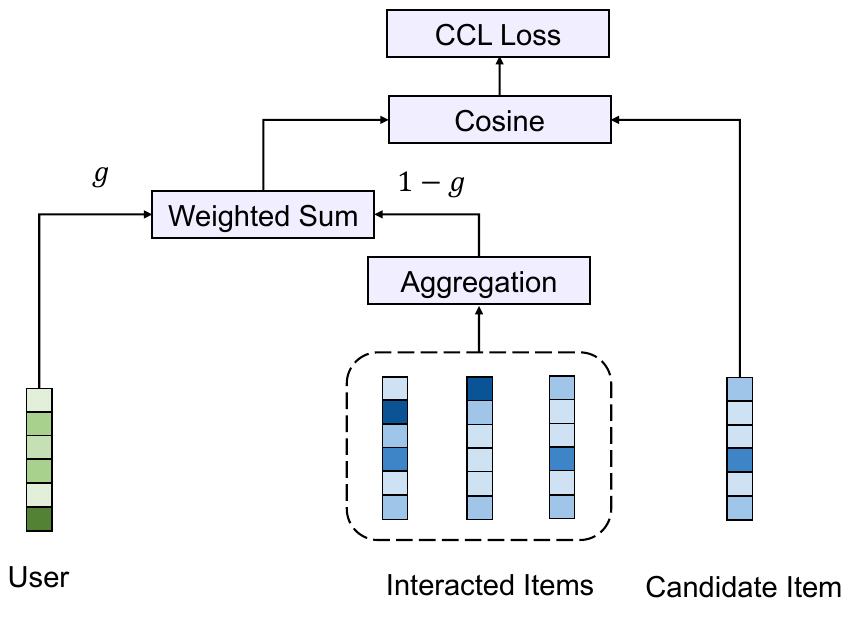}
    \end{center}
    % \vspace{-2ex}
    \caption{The Simple Model Architecture of SimpleX.}
    \label{fig:overview}
\end{figure}
%%%%%%%%%%%%%%%%%%%%%%%%%%%%%%%%%%%

\subsection{Model Architecture}
\label{simplecf}
To leverage the advantages of CCL, we further propose a simple CF model, dubbed SimpleX. In the design of SimpleX, we keep simplicity in mind and borrow ideas from several successful models such as YouTubeNet~\cite{YouTubeNet}, ACF~\cite{ACF}, and PinSage~\cite{PinSage}.

Figure~\ref{fig:overview} illustrates the overall architecture of SimpleX. It largely follows the mechanism of MF, which factorizes users and items into a common latent space. Yet, SimpleX also takes the interacted item sequence of each user as additional input to better model user behaviors. This also has been shown effective in many existing studies, such as YouTubeNet~\cite{YouTubeNet} and ACF~\cite{ACF}. The key part of SimpleX lies in its aggregation layer for behavior sequence aggregation. 
Here we introduce three common aggregation choices, including average pooling, self-attention, and user-attention, but Simplex is a unified architecture that any other aggregation method should also be applicable.

Suppose the historically interacted item set of user $u$ as $\mathcal{H}_u$, and we set its maximal size to $K$. For users with a different size of interacted items, either padding or chunking can be applied accordingly. As such, the aggregated vector can be obtained as follows:
\begin{equation}
{p}_u = \sum\nolimits_{k \in \mathcal{H}_u} I_k\cdot \alpha_k \boldsymbol{e}_k~, \label{equ_aggregation}
\end{equation}
where $e_k\in \mathbb{R}^{d \times 1}$ is the $d$-dimensional embedding vector of item $k$. $I_k$ denotes the mask indices to $\mathcal{H}_u$ during padding, where $I_k=0$ indicates a padding token; otherwise $I_k=1$. $\alpha_k$ denotes the aggregation weight, which can be computed according to different aggregation types as follows.
\begin{equation}
   \alpha_k = \begin{cases}
   \frac{I_k}{\sum\nolimits_{k\in \mathcal{H}_u}I_k}~, & \text{if average pooling}, \\[6pt]
    \frac{I_k\cdot exp(\beta_k)}{\sum_{j \in \mathcal{H}_u} I_k\cdot exp(\beta_j)}~, & \text{if self-attention or user-attention}.
 \end{cases}\label{attention_weight}
\end{equation}

Average pooling provides a straightforward way to aggregate the interacted items, which has been successfully applied in YouTubeNet~\cite{YouTubeNet}. But it treats each item equally and fails to account for the relative importances of different items as well as a user's preference on each item. The attention mechanism, such as self-attention and user-attention, can be applied in such cases as calculated in the lower part of Equation~\ref{attention_weight}. The difference between them lies in the computation of $\beta_k$, which is:
\begin{equation}
   \beta_k = \begin{cases}
   {q}^Ttanh(\boldsymbol{W}_1{e}_k + {b}_1)~, & \text{if self-attention}, \\[2pt]
    {e}_u^Ttanh(\boldsymbol{W}_2 {e}_k + {b}_2)~, & \text{if user-attention}.
 \end{cases}
\end{equation}
where $q\in \mathbb{R}^{d \times 1}$ is a learnable global query vector for self-attention and ${e}_u$ is the user-specific query vector for user $u$ in user-attention. ${\boldsymbol{W}_1},{\boldsymbol{W}_2} \in  \mathbb{R}^{d\times d}$ and ${b}_1,{b}_2 \in \mathbb{R}^{d\times1}$ are learnable parameters. Note that similar attention mechanisms can be found in some existing work~\cite{ACF,NPA}.

%While YouTubeNet is effective in industrial recommender systems~\cite{YouTubeNet}, it only employs average pooling to aggregate the historical behavior sequence of each user. Inspired by the widespread use of attention mechanisms in recommender systems (e.g.,~\cite{DIN,NPA}), we propose a simple baseline collaborative filtering approach by leveraging the attention mechanism. Given a user vector and a sequence of interacted items, we design two attention modules to aggregate the item sequence, including self-attention and cross-attention. While self-attention aims to aggregate item sequence from a global view, cross-attention is used to aggregate item sequence from a personalized perspective.

However, after behavior aggregation via Equation~\ref{equ_aggregation}, the pooling vector $p_u$ may lie in a different latent space with user vector $e_u$. We further fuse both parts to get the final user representation ${h}_u$:
\begin{equation}
{h}_u = g\cdot {e}_u + (1-g)\cdot V{p}_u
\end{equation}
where $\boldsymbol{V} \in  \mathbb{R}^{d\times d}$ is a learnable parameter and $g$ is a hyperparameter weight. % which can be either learned end-to-end or set as a hyper-parameter to tune. 
Finally, we measure the cosine similarity $\hat{y}_ui$ between user $u$ and item $i$ as the input to our CCL loss.
\begin{equation}
\hat{y}_{ui} = cos(h_u, e_i)
\end{equation}
% Especially, when both $h_u$ and $e_i$ are normalized to unit vectors, it equals to the dot similarity that is commonly used in MF. 
The above three aggregation layers provide different views for aggregation, including global-average view, global-weighed view and user-specific weighted view. The choice among them is quite data-dependent. In our experiment, we show that average-pooling is a robust aggregation method that always demands a first attempt when applying SimpleX. The other two usually needs more efforts to tune and in some cases brings marginal improvements. 

\subsection{Relationships to Existing Models}
SimpleX is also related to multiple popular CF models. 
\begin{itemize}
    \item \textbf{MF}. MF is the most common model for CF. SimpleX follows the similar mechanism of MF. When setting g = 1 in SimpleX, it reduces to a MF model trained with CCL (i.e., MF-CCL).
    
    \item \textbf{YouTubeNet}. YouTubeNet is a successful model that has been widely used in industry. SimpleX can be also seen as a simplified YouTubeNet model (without using side features) when average pooling is employed. The only difference is that YouTubeNet employs concatenation instead of weighted sum to fuse $e_u$ and $p_u$. But the latter performs better in our experiments.
    
    \item \textbf{GNN-based models}. Simplex is also similar to GNN-based models. For instance, when choosing the user-attention aggregation layer, it almost equals to a graph attention (GAT) layer applied on user nodes only. If using the self-attention aggregation layer, it works like the neighbor interaction in NIA-GCN~\cite{NIA-GCN} as well.  
\end{itemize}

We emphasize that although the design of SimpleX is simple and might not be novel to some extent, it unifies several key components in existing CF models. \textbf{Surprisingly, such a simple model is sufficient to surpass most state-of-the-art CF models by a large margin, which could serve as simple and strong baseline for future research}.

%% file: sections/4_experiment.tex
\begin{table*}[!t]
\centering
\caption{Performance of MF under different loss functions. The best result in each column is marked in bold.}
\begin{tabular}{c|cc|cc|cc}
\hline
\multirow{2}{*}{Loss} & \multicolumn{2}{c|}{AmazonBooks}  & \multicolumn{2}{c|}{Yelp18}       & \multicolumn{2}{c}{Gowalla}       \\ \cline{2-7} 
                    & Recall@20 & NDCG@20 & Recall@20 & NDCG@20 & Recall@20 & NDCG@20 \\ \hline\hline
BPR Loss             & 0.0338    & 0.0261  & 0.0549    & 0.0445  & 0.1616    & 0.1366  \\
Pairwise Hinge Loss   & 0.0352    & 0.0267  & 0.0562    & 0.0453  & 0.1318    & 0.0996  \\
Binary Cross-Entropy & 0.0479   & 0.0371  & 0.0617    & 0.0503  & 0.1321    & 0.1159  \\
Softmax Cross-Entropy & 0.0478    & 0.0367  & 0.0639    & 0.0522  & 0.1545    & 0.1276  \\
Mean Square Error     & 0.0337    & 0.0267  & 0.0624    & 0.0513  & 0.1528    & 0.1315  \\
Cosine Contrastive Loss   & \textbf{0.0559} & \textbf{0.0447} & \textbf{0.0698} & \textbf{0.0572} & \textbf{0.1837} & \textbf{0.1493} \\ \hline
\end{tabular}
\label{DifferentLoss}

\end{table*}
%%%%%%%%%%%%%%%%%%%%%%%%

\section{Experiments}\label{sec:exp}
% In this section, we conduct extensive experiments to evaluate our proposal. 
% We first describe experimental settings (section~\ref{}), and then analyze the effect of different loss functions and negative sampling ratios in section~\ref{}. 
% We next compare our three proposed baselines with other state-of-the-art methods in section~\ref{}.
% To justify the design of SimpleCF, we perform ablation studies on SimpleCF in section~\ref{}.
% We also show how LightGCN, which is the most powerful CF model now, will perform with new loss functions to get more insight, in section~\ref{}. 
% Finally, we compare the efficiency of our three baselines with others in section~\ref{}.

% Our experiments are aim to answer the following research questions:
% \begin{itemize}
%     \item \textbf{RQ1:} Are our concerns about the default evaluation setting in current CF research justified? What are the effects of loss functions and negative sampling ratios on CF?
%     \item \textbf{RQ2:} Can our MF-CCL and SimpleX be strong baselines for CF research? What are their performances compared with different types of SOTA methods on various datasets?
%     \item \textbf{RQ3:} How does other models perform when training with CCL? Is CCL promising to become a widely-used loss function in the future CF studies? 
%     \item \textbf{RQ4:} How the different neighbor pooling strategies and combination weights of SimpleX affects its performance?
%     \item \textbf{RQ5:} How much advantage does our proposed baseline models have versus the SOTA models in terms of efficiency?
% \end{itemize}
In this section, we conduct comprehensive experiments to evaluate SimpleX, including: 1) studying the impacts of loss functions and negative sampling ratios, 2) making performance comparisons to existing models on three main datasets, 3) incorporating CCL to other models, 4) performing parameter analysis and efficiency evaluation, 5) further validating SimpleX on some other datasets.
 
\subsection{Experimental Setup}
\subsubsection{Dataset}
\label{Dataset}
We use 11 benchmark datasets in our study. For fairness and ease of comparison, we choose those open datasets that have been already split and preprocessed. Specifically:

(1) We employ three main datasets \textbf{Amazon-Books}, \textbf{Yelp2018}, and \textbf{Gowalla}, which are commonly used in recent GNN-based CF models~\cite{NGCF, LR-GCCF, LightGCN, DGCF, NGAT4Rec, SGL-ED}. We perform most of our experiments on them and further make comparisons to these GNN-based models. 

    (2) To demonstrate the universality of SimpleX, we further test SimpleX on some other datasets adopted by studies published in top-tier conferences. Three of them, \textbf{Amazon-CDs}, \textbf{Amazon-Movies}, \textbf{Amazon-Beauty}, are adopted by the work NIA-GCN~\cite{NIA-GCN} and BGCF~\cite{BGCF}. The other three, \textbf{Amazon-Electronics}, \textbf{CiteUlike-A}, and \textbf{Movielens-1M}, are provided by NBPO~\cite{NBPO}, DHCF~\cite{DHCF}, and LCFN~\cite{LCFN}, respectively. 
Specifically, we compare SimpleX with the corresponding models on the corresponding datasets that adopted in their original papers. For example, we will compare with DHCF~\cite{DHCF} on CiteUlike-A dataset because DHCF adopts this dataset in their original paper.

(3) The last two are \textbf{Movielens-20M} and \textbf{MillionSongData}, which are commonly used by autoencoder-based CF models, such as Mult-VAE~\cite{MVAE} and RecVAE~\cite{RecVAE}.
We follow the strong generalization setting, which split train/validation/test sets with different sets of users, and specially make comparison with those autoencoder-based CF models to further demonstrate the effectiveness of SimpleX.

% We use this dataset to demonstrate the 
% More details of these datasets can be found in the corresponding papers. We follow the exactly same settings with them.

%Early stopping will be triggered if recall@20 does not increase for 3 successive epochs.  

% To evaluate the performance of proposed methods, we use three publicly available datasets: Gowalla, Yelp2018 and Amazon-Book in the experiments, which are also used in other studies ~\cite{NGCF,LightGCN,DGCF}. The statistics of the them are shown in Table~\ref{data_statistics}. The detail information and the method of data split can be found in the work NGCF ~\cite{NGCF}. We employ NDCG@K and Recall@K as the evaluation metrics, which are widely used in recommender system studies, and we set K as to 20.
%We choose three publicly available datasets: Gowalla, Yelp2018, and Amazon-Book, which are also used in the experiments of other state-of-the-art models~\cite{NGCF,LightGCN,DGCF}, to validate our proposals. The statistics of datasets are concluded in Table~\ref{data_statistics}, and we follow the same data split as NGCF. We choose recall@20 and ndcg@20, which are widely used in CF research, as the evaluation metrics.

\subsubsection{Compared Methods}
We compare SimpleX with 29 existing CF models of different types:
\begin{itemize}
    \item  Five methods based on MF and its variants, including MF-BPR~\cite{MF}, GRMF~\cite{GRMF}, HOP-Rec~\cite{HOP-Rec}, NeuMF~\cite{NeuMF}, and ENMF~\cite{ENMF};
    \item Four autoencoder-based methods, including Mult-VAE~\cite{MVAE}, MacridVAE~\cite{MacridVAE}, EASE\textsuperscript{R}~\cite{EASE}, and RecVAE~\cite{RecVAE};
    \item Fourteen GNN-based methods, including GC-MC~\cite{GC-MC}, Pinsage~\citep{PinSage}, GAT~\cite{GAT}, NGCF~\cite{NGCF}, DisenGCN~\cite{DisenGCN}, LR-GCCF~\cite{LR-GCCF}, NIA-GCN~\cite{NIA-GCN}, LightGCN~\cite{LightGCN}, DGCF~\cite{DGCF}, NGAT4Rec~\cite{NGAT4Rec}, SGL-ED~\cite{SGL-ED}, BGCF~\cite{BGCF}, DHCF~\cite{DHCF}, and LCFN~\cite{LCFN};
    \item Six methods of other types, including ItemPop, SLIM~\cite{SLIM}, CML~\cite{CML}, YouTubeNet~\cite{YouTubeNet},  CMN~\cite{CMN}, and NBPO~\cite{NBPO}.
\end{itemize}
% (1) Five methods based on MF and its variants, including MF-BPR~\cite{MF}, GRMF~\cite{GRMF}, HOP-Rec~\cite{HOP-Rec}, NeuMF~\cite{NeuMF}, and ENMF~\cite{ENMF}; (2) Three autoencoder-based methods, including Mult-VAE~\cite{MVAE}, MacridVAE~\cite{MacridVAE}, and EASE\textsuperscript{R}~\cite{EASE}; (3) Fourteen GNN-based methods, including GC-MC~\cite{GC-MC}, Pinsage~\citep{PinSage}, GAT~\cite{GAT}, NGCF~\cite{NGCF}, DisenGCN~\cite{DisenGCN}, LR-GCCF~\cite{LR-GCCF}, NIA-GCN~\cite{NIA-GCN}, LightGCN~\cite{LightGCN}, DGCF~\cite{DGCF}, NGAT4Rec~\cite{NGAT4Rec}, SGL-ED~\cite{SGL-ED}, BGCF~\cite{BGCF}, DHCF~\cite{DHCF}, and LCFN~\cite{LCFN}; (4) Six methods of other types, including ItemPop, SLIM~\cite{SLIM}, CML~\cite{CML}, YouTubeNet~\cite{YouTubeNet},  CMN~\cite{CMN}, and NBPO~\cite{NBPO}.

% We emphasize that all models compared in the same table  (we split the experiments to three parts based on the datasets (see Section~\ref{performance_comparison_to_sota} ))
% Note that when comparing SimpleX with the baselines, \textbf{we ensure all of the compared baselines are evaluated under the exactly same evaluation protocol.}

\subsubsection{Implementation Details}
\label{implementation}
We implement SimpleX in PyTorch. Specifically, we set the batch size to 1024 by default. We use the Adam optimizer and tune the learning rate among [1e-3, 5e-4, 1e-4]. We also employ $L_2$ regularization on the embedding parameters and search the regularization weight between 1e-9$\sim$1e-2 with an increase ratio of 5. For cosine contrastive loss, we search the number of negative samples from 1 to 2000. In many cases, we pick 100, 500, or 1000. The margin $m$ is tuned among 0$\sim$1 at an interval of 0.1, for example, we set 0.4, 0.9, and 0.9 on Amazon-Books, Yelp2018, and Gowalla, respectively. %We set the maximal size of interacted items $K = 500$. 
Meanwhile, we use the same embedding size with the compared model, for example, 64 in LightGCN and 128 in LCFN. For fairness of comparison with existing models, we reuse the reported results from their papers and compare with SimpleX using the same evaluation metrics (e.g., Recall@20, NDCG@20, or F1@20) for consistency. To facilitate reproducible research in the community, we have open our source code and detailed benchmark settings at RecZoo.

% For all compared models and SimpleX, hyper-parameters are carefully tuned based on the validation set and the same proper early stopping strategy is used to prevent overfitting. 
% Due to the space limitation, more detailed settings for each model can be found in our open code later.

% For most of our compared models, we directly copy the results from their papers, since we follow the same experimental settings. This also ensures fairness of comparison since the published results from the authors are usually well tuned. In addition, we add some other remarkable models, including SLIM, CML, and YouTubeNet. We use the open-source code from Github and carefully tune the hyperparameters. We will release our source code along with the detailed parameter settings for reproduction. 

\subsection{Impact of Different Loss Functions}\label{subsec:lossFunc}
While most studies focus on the interaction encoder design, they neglect the importance of loss functions in the learning of a CF model. We make a systematic comparison on the impacts of different loss functions. For this purpose, we choose one of the simplest baseline CF models, i.e., MF, as the backbone to perform the experiments, since simple models tend to be more illustrative. In addition to our CCL loss, we evaluate MF on the following representative loss functions:
\begin{itemize}
    \item \textbf{Bayesian personalized ranking (BPR)} loss 
    encourages the similarity of a positive user-item pair to be higher than that of each negative user-item pair. It is one of the most commonly used loss function for CF research~\cite{BPR,NGCF,LightGCN,DGCF}.

    \item \textbf{Pairwise hinge loss (PHL)}, is also known as max-margin objective, which has been used in CML~\citep{CML}. PHL forces the distance of a negative user-item pair to be larger than a positive one by at least the marginal distance.
    
    \item \textbf{Binary cross-entropy (BCE)} loss is commonly used for binary classification, which has been adopted in the early work NeuMF~\cite{NeuMF}.
    
    \item \textbf{Softmax cross-entropy (SCE)} loss is widely used for multi-class classification. YouTubeNet~\cite{YouTubeNet} cast item prediction as a multi-class classification task through the SCE loss.

    %  \item \textbf{NonSamplingLoss (NSL)} is the loss function used in ENMF~\cite{ENMF}. Essentially, this loss is weighted Mean Square Error (MSE) loss optimized in an more efficient way which can make use of all missing data.
     
      \item \textbf{Mean square error (MSE)} has been widely used for CF, such as WMF~\cite{WMF} and ENMF~\cite{ENMF}. 
 
\end{itemize}

Table~\ref{DifferentLoss} shows the results of training MF with different loss functions on Amazon-Books, Yelp2018, and Gowalla. Note that every model has been trained with enough epochs to reach convergence and the best results are reported. From the results, we have the following observations:

1) CCL consistently achieves the best performance on all the three datasets, outperforming the other loss functions by at least 16.7\%, 9.2\% and 13.7\% w.r.t. Recall@20 on Amazon-Books, Yelp2018 and Gowalla, respectively.
    
2) BPR only appears to be strong on Gowalla and performs not well on both Amazon-Books and Yelp2018. This demonstrates that using BPR for training is probably sub-optimal, and thus the results reported by many previous papers may need careful re-examination and are likely to be further improved with our CCL loss.

\textbf{Why CCL performs better than the other loss functions?} 
In addition to the design choices analyzed in Section~\ref{approach:MF-CCL}, we further highlight the advantages of CCL with some concrete comparisons.
% We attribute the excellent performance of CCL to the following reasons. 
%First, CCL is a pointwise loss function, which does not require the number of positive and negative samples to be the same. This provides us a change to adjust the proportion of losses between positive samples and negative samples to be in accord with the dataset distribution so as to achieve better performance. 
First, in contrast to BPR, BCE, SCE, and MSE, CCL can automatically filter out hard negative samples that are hard to distinguish (i.e., large cosine similarity) by the model via its margin mechanism. For example, if we set $m = 0.8$, only those negative pairs with $\hat{y}_{ui} > 0.8$ will contribute to the loss. Different from the above loss functions that treat each negative sample equally, CCL allows the model to emphasize on the learning of hard negative samples and thus generate more discriminative representations.
Second, compared with PHL that also applies a margin mechanism, CCL is more effective for CF. The PHL loss is determined by the relative distance between positive samples and negative samples. Even if a negative sample is actually hard to be distinguished (e.g., $\hat{y}^{neg}_{ui}=0.8$), it will not contribute to learning if the corresponding positive sample has $\hat{y}^{pos}_{ui}=0.9$. CCL avoids such ambiguity by penalizing the absolute similarity $\hat{y}_{ui}$ of each negative sample.

% Third, as implicit CF is more like a classification task but not regression task, 

% compared with MSE which tries to reduce the Euclidean distance between the matching score of the user-item pair and a fixed value (e.g., 1 for positive pairs and 0 for negative pairs)

% This may be due to the good combination of the pointwise property and the absolute margin used in CCL, which are suitable for CF as we illustrated in section~\ref{Prolossfunc}, and hence we recommend considering CCL as a regular loss function for future CF research.

\subsection{Impact of Negative Sampling Ratio}\label{subsec:negSampRatio}

\begin{figure}[!t]
    \begin{center}
    %\framebox[4.0in]{$\;$}
    % \fbox{\rule[-.5cm]{0cm}{4cm} \rule[-.5cm]{4cm}{0cm}}
    \includegraphics[width = 0.95\linewidth]{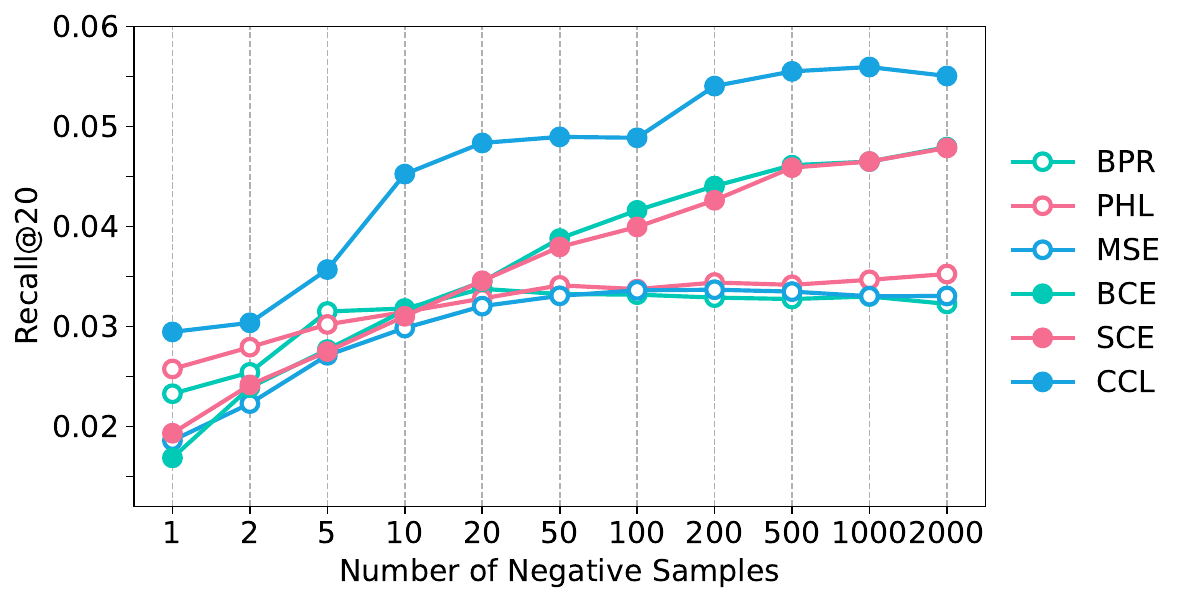}
    \end{center}
    \vspace{-2ex}
    \caption{The effect of number of negative samples across different loss functions on Amazon-Books.}
    \label{fig:num_negs}
    \vspace{-0.5cm}
\end{figure}

%%%%%%%%%%%%%%%%%%%%%%%%%%%%%%%%%%%%%%%%%%%%
%
%	\centering  %图片全局居中
%	\subfigure[Recall]{
%		
%		\includegraphics[width=0.3\columnwidth]{fig/LightGCN-amazon.pdf}}
%	\subfigure[NDCG]{
%		
%		\includegraphics[width=0.3\columnwidth]{fig/LightGCN-amazon.pdf}}
%	\subfigure[Recall]{
%		
%		\includegraphics[width=0.3\columnwidth]{fig/LightGCN-amazon.pdf}}
%	\subfigure[Recall]{
%		
%		\includegraphics[width=0.3\columnwidth]{fig/LightGCN-amazon.pdf}}
%	\subfigure[Recall]{
		
%		\includegraphics[width=0.3\columnwidth]{fig/LightGCN-amazon.pdf}}
%	\subfigure[Recall]{
%		
%		\includegraphics[width=0.3\columnwidth]{fig/LightGCN-amazon.pdf}}
%	\caption{Performance comparison among variants of UltraGCN and LightGCN on Amazon-Books.}
%	\label{fig_ablation_study}
%\end{figure*}
%

We argue that negative sampling ratio is also important in the learning of CF models, which has been largely ignored by existing studies. To support our claims, we compare the performance of MF trained with 1$\sim$2000 negative samples on Amazon-Books. We also repeat the experiment on different loss functions. We train each model until convergence and report the best results, as shown in Figure~\ref{fig:num_negs}. We have the following observations from the results:

1) The number of negative samples does matter for CF model training. Generally, increasing it within a certain range leads to improvements. This suggests that we should carefully consider the impact of the number of negative samples in the evaluation. 

2) MF trained with CCL is consistently better than training with the other loss functions under different negative sampling ratios, further demonstrating the superiority of our CCL.
    
3) The performances of PHL, MSE, and BPR become stable when the number of negative samples increases to 50. In contrast, CCL, BCE, and SCE can keep performance gains with the increase of number of negative samples, even when it reaches to 1000.

In summary, our experimental results show that both loss functions and negative sampling ratios can have a large impact on model performance. Training with the CCL loss and a large negative sampling ratio appears to be a promising setting for CF methods to gain higher performance. We therefore call for more future research towards this direction. 

% The prevailing setting that only using BPR with one negative sample is flawed, and thus the current baselines are not strong enough for the benign development of the CF community. 

%We would like to suggest that researchers should take experiments with more collocations of loss functions and negative sampling ratios before proposing their new CF models.

%%%%%%%%%%%%%%%%%%%%%%%%%%%%%%%%%%%%%%%%%%%%%%%%%
\begin{table*}[!t]
\renewcommand\arraystretch{1.01}
%\small
\centering
\caption{Performance comparison to popular CF models. We highlight the top-5 best results in each column. RI stands for relative improvement. We also report the average RI over NGCF for each model if the improvement is positive. For models marked by $^\ast$, the results are reported by the existing papers and we reuse them for consistency.}
\begin{tabular}{rcllllllll}
\hline
\multicolumn{1}{r|}{} &
  \multicolumn{1}{c|}{} &
  \multicolumn{2}{c|}{Amazon-Books} &
  \multicolumn{2}{c|}{Yelp2018} &
  \multicolumn{2}{c|}{Gowalla} &
  \multicolumn{2}{c}{Avg RI over NGCF} \\ \cline{3-10} 
\multicolumn{1}{r|}{\multirow{-2}{*}{Publication}} &
  \multicolumn{1}{c|}{\multirow{-2}{*}{Model}} &
  \hspace{-0.5ex}Recall@20 &
  \multicolumn{1}{l|}{\hspace{-0.5ex}NDCG@20} &
  \hspace{-0.5ex}Recall@20 &
  \multicolumn{1}{l|}{\hspace{-0.5ex}NDCG@20} &
  \hspace{-0.5ex}Recall@20 &
  \multicolumn{1}{l|}{\hspace{-0.5ex}NDCG@20} &
  \hspace{-0.5ex}Recall@20 &
  \hspace{-0.5ex}NDCG@20 \\ \hline
\multicolumn{1}{r|}{--} &
  \multicolumn{1}{c|}{ItemPop} &
  0.0051 &
  \multicolumn{1}{l|}{0.0044} &
  0.0124 &
  \multicolumn{1}{l|}{0.0101} &
  0.0416 &
  \multicolumn{1}{l|}{0.0317} &
  -- &
  -- \\
  \multicolumn{1}{r|}{UAI'2009} &
  \multicolumn{1}{c|}{MF-BPR} &
  0.0338 &
  \multicolumn{1}{l|}{0.0261} &
  0.0576 &
  \multicolumn{1}{l|}{0.0468} &
  0.1627 &
  \multicolumn{1}{l|}{0.1378} &
  -- &
  -- \\
% \multicolumn{1}{r|}{ICDM'2011} &
%   \multicolumn{1}{c|}{SLIM} &
%   0.0470 &
%   \multicolumn{1}{l|}{0.0377} &
%   0.0646  &
%   \multicolumn{1}{l|}{\textbf{0.0541(5)}} &
%  0.1699  &
%   \multicolumn{1}{l|}{0.1382} & 
% 24.1\% &
% 28.4\% \\
\multicolumn{1}{r|}{NIPS'2015} &
  \multicolumn{1}{c|}{GRMF$^\ast$} &
  0.0354 &
  \multicolumn{1}{l|}{0.0270} &
  0.0571 &
  \multicolumn{1}{l|}{0.0462} &
  0.1477 &
  \multicolumn{1}{l|}{0.1205} &
  -- &
  -- \\
\multicolumn{1}{r|}{RecSys'2016} & 

  \multicolumn{1}{c|}{YouTubeNet} &
  \textbf{0.0502(4)} &
  \multicolumn{1}{l|}{\textbf{0.0388(4)}} &
  \textbf{0.0686(3)} &
  \multicolumn{1}{l|}{\textbf{0.0567(3)}} &
  \textbf{0.1754(5)} &
  \multicolumn{1}{l|}{\textbf{0.1473(5)}} &
  32.2\% &
  33.3\% \\
\multicolumn{1}{r|}{WWW'2017} &
  \multicolumn{1}{c|}{NeuMF$^\ast$} &
  0.0258 &
  \multicolumn{1}{l|}{0.0200} &
  0.0451 &
  \multicolumn{1}{l|}{0.0363} &
  0.1399 &
  \multicolumn{1}{l|}{0.1212} &
  -- &
  -- \\
\multicolumn{1}{r|}{WWW'2017} &
  \multicolumn{1}{c|}{CML} &
  \textbf{0.0522(3)} &
  \multicolumn{1}{l|}{\textbf{0.0428(3)}} &
  0.0622 &
  \multicolumn{1}{l|}{0.0536} &
  0.1670 &
  \multicolumn{1}{l|}{0.1292} &
  29.6\% &
  37.6\% \\
\multicolumn{1}{r|}{SIGIR'2018} &
  \multicolumn{1}{c|}{CMN$^\ast$} &
  0.0267 &
  \multicolumn{1}{l|}{0.0218} &
  0.0457 &
  \multicolumn{1}{l|}{0.0369} &
  0.1405 &
  \multicolumn{1}{l|}{0.1221} &
  -- &
  -- \\
\multicolumn{1}{r|}{RecSys'2018} &
  \multicolumn{1}{c|}{HOP-Rec$^\ast$} &
  0.0309 &
  \multicolumn{1}{l|}{0.0232} &
  0.0517 &
  \multicolumn{1}{l|}{0.0428} &
  0.1399 &
  \multicolumn{1}{l|}{0.1214} &
  -- &
  -- \\
\multicolumn{1}{r|}{WWW'2018} &
  \multicolumn{1}{c|}{Mult-VAE$^\ast$} &
  0.0407 &
  \multicolumn{1}{l|}{0.0315} &
  0.0584 &
  \multicolumn{1}{l|}{0.0450} &
  0.1641 &
  \multicolumn{1}{l|}{0.1335} &
  9.6\% &
  7.1\% \\
\multicolumn{1}{r|}{NeurIPS'2019} &
  \multicolumn{1}{c|}{MacridVAE$^\ast$} &
  0.0383 &
  \multicolumn{1}{l|}{0.0295} &
  0.0612 &
  \multicolumn{1}{l|}{0.0495} &
  0.1618 &
  \multicolumn{1}{l|}{0.1202} &
  8.5\% &
  8.0\% \\
\multicolumn{1}{r|}{TOIS'2020} &
  \multicolumn{1}{c|}{ENMF} &
  0.0359 &
  \multicolumn{1}{l|}{0.0281} &
  0.0624 &
  \multicolumn{1}{l|}{0.0515} &
  0.1523 &
  \multicolumn{1}{l|}{0.1315} &
  6.1\% &
  7.4\% \\ \hline
\multicolumn{10}{c}{\cellcolor[HTML]{F8F8F8}\textbf{GNN-based Models}} \\ \hline
\multicolumn{1}{r|}{KDDW'2018} &
  \multicolumn{1}{c|}{GC-MC$^\ast$} &
  0.0288 &
  \multicolumn{1}{l|}{0.0224} &
  0.0462 &
  \multicolumn{1}{l|}{0.0379} &
  0.1395 &
  \multicolumn{1}{l|}{0.1204} &
  -- &
  -- \\
\multicolumn{1}{r|}{KDD'2018} &
  \multicolumn{1}{c|}{PinSage$^\ast$} &
  0.0282 &
  \multicolumn{1}{l|}{0.0219} &
  0.0471 &
  \multicolumn{1}{l|}{0.0393} &
  0.1380 &
  \multicolumn{1}{l|}{0.1196} &
  -- &
  -- \\
\multicolumn{1}{r|}{ICLR'2018} &
  \multicolumn{1}{c|}{GAT$^\ast$} &
  0.0326 &
  \multicolumn{1}{l|}{0.0235} &
  0.0543 &
  \multicolumn{1}{l|}{0.0431} &
  0.1401 &
  \multicolumn{1}{l|}{0.1236} &
  -- &
  -- \\
\multicolumn{1}{r|}{SIGIR'2019} &
  \multicolumn{1}{c|}{NGCF$^\ast$} &
  0.0344 &
  \multicolumn{1}{l|}{0.0263} &
  0.0579 &
  \multicolumn{1}{l|}{0.0477} &
  0.1570 &
  \multicolumn{1}{l|}{0.1327} &
  -- &
  -- \\
\multicolumn{1}{r|}{ICML'2019} &
  \multicolumn{1}{c|}{DisenGCN$^\ast$} &
  0.0329 &
  \multicolumn{1}{l|}{0.0254} &
  0.0558 &
  \multicolumn{1}{l|}{0.0454} &
  0.1356 &
  \multicolumn{1}{l|}{0.1174} &
  -- &
  -- \\
\multicolumn{1}{r|}{AAAI'2020} &      

  \multicolumn{1}{c|}{LR-GCCF} &
  0.0335 &
  \multicolumn{1}{l|}{0.0265} &
  0.0561 &
  \multicolumn{1}{l|}{0.0343} &
  0.1519 &
  \multicolumn{1}{l|}{0.1285} &
  -- &
 -- \\
\multicolumn{1}{r|}{SIGIR'2020} &
  \multicolumn{1}{c|}{NIA-GCN$^\ast$} &
  0.0369 &
  \multicolumn{1}{l|}{0.0287} &
  0.0599 &
  \multicolumn{1}{l|}{0.0491} &
  0.1359 &
  \multicolumn{1}{l|}{0.1106} &
  6.9\% &
  4.8\% \\
\multicolumn{1}{r|}{SIGIR'2020} &
  \multicolumn{1}{c|}{LightGCN$^\ast$} &
  0.0411 &
  \multicolumn{1}{l|}{0.0315} &
  0.0649 &
  \multicolumn{1}{l|}{0.0530} &
  \textbf{0.1830(4)} &
  \multicolumn{1}{l|}{\textbf{0.1554(3)}} &
  15.8\% &
  15.4\% \\
\multicolumn{1}{r|}{SIGIR'2020} &
  \multicolumn{1}{c|}{DGCF$^\ast$} &
  0.0422 &
  \multicolumn{1}{l|}{0.0324} &
  0.0654 &
  \multicolumn{1}{l|}{0.0534} &
  \textbf{0.1842(2)} &
  \multicolumn{1}{l|}{\textbf{0.1561(1)}} &
  17.8\% &
  17.6\% \\
\multicolumn{1}{r|}{Arxiv'2020} &
  \multicolumn{1}{c|}{NGAT4Rec$^\ast$} &
  0.0457 &
  \multicolumn{1}{l|}{0.0358} &
  \textbf{0.0675(4)} &
  \multicolumn{1}{l|}{\textbf{0.0554(5)}} &
  -- &
  \multicolumn{1}{l|}{--} &
  24.7\% &
  26.1\% \\
\multicolumn{1}{r|}{SIGIR'2021} &
  \multicolumn{1}{c|}{SGL-ED$^\ast$} &
  \textbf{0.0478(5)} &
  \multicolumn{1}{l|}{\textbf{0.0379(5)}} &
  \textbf{0.0675(4)} &
  \multicolumn{1}{l|}{\textbf{0.0555(4)}} &
  -- &
  \multicolumn{1}{l|}{--} &
  27.8\% &
  30.2\% \\
 \hline
\multicolumn{10}{c}{\cellcolor[HTML]{F8F8F8}\textbf{Ours}} \\ \hline
\multicolumn{1}{r|}{CIKM'2021} &
  \multicolumn{1}{c|}{MF-CCL} &    
  \textbf{0.0559(2)} &
  \multicolumn{1}{l|}{\textbf{0.0447(2)}} &
  \textbf{0.0698(2)} &
  \multicolumn{1}{l|}{\textbf{0.0572(2)}} &
  \textbf{0.1837(3)} &
  \multicolumn{1}{l|}{\textbf{0.1493(4)}} &
  41.6\% &
  45.0\% \\
\multicolumn{1}{r|}{CIKM'2021} &  
  \multicolumn{1}{c|}{SimpleX} &
  \textbf{0.0583(1)} &
  \multicolumn{1}{l|}{\textbf{0.0468(1)}} &
  \textbf{0.0701(1)} &
  \multicolumn{1}{l|}{\textbf{0.0575(1)}} &
  \textbf{0.1872(1)} &
  \multicolumn{1}{l|}{\textbf{0.1557(2)}} &
  45.3\% &
  49.2\% \\ \hline
\multicolumn{2}{c|}{RI over NGCF} &
  69.6\% &
  \multicolumn{1}{l|}{77.9\%} &
  21.1\% &
  \multicolumn{1}{l|}{20.6\%} &
  19.2\% &
  \multicolumn{1}{l|}{17.3\%} &
   &
   \\
\multicolumn{2}{c|}{RI over LighGCN} &
  41.9\% &
  \multicolumn{1}{l|}{48.5\%} &
  8.0\% &
  \multicolumn{1}{l|}{8.5\%} &
  2.3\% &
  \multicolumn{1}{l|}{0.2\%} &
   &
   \\ \hline
\end{tabular}
% \vspace{0.5ex}
%     \begin{flushleft}{\hspace{2ex} $^*$ Note that the superscript marked on each model indicates where the results are copied from: a~\cite{NGCF}, b~\cite{LightGCN}, c~\cite{DGCF}, d~\cite{NGAT4Rec}, e~\cite{SGL-ED} }\end{flushleft}
\label{fullresults1}
\end{table*}

\subsection{Performance Comparison to SOTA Models}
\label{performance_comparison_to_sota}
In this section, we provide a comprehensive comparison results of SimpleX and other 23 CF models on three main datasets, i.e., Amazon-Books, Yelp2018, and Gowalla, which are very commonly adopted in CF studies (especially in GNN-based CF), to demonstrate the superiority of SimpleX. 
Table~\ref{fullresults1} shows our performance comparisons on Amazon-Books, Yelp2018, and Gowalla under the same evaluation protocol, and we have the following observations:

% We compare SimpleX with a total of 28 existing CF models to demonstrate its superiority and generability as a potential strong baseline for future research. We divide these comparisons into three parts. 

1) Our SimpleX achieves the best overall performance on all the three datasets. In particular, compared with the most recent LightGCN, SimpleX makes 41.9\%, 8.0\%, and 2.3\% performance improvements on Recall@20 for Amazon-Books, Yelp2018, and Gowalla, respectively, demonstrating the high effectiveness of SimpleX. Besides, note that we do not report the results of SGL-ED~\cite{SGL-ED} and NGAT4Rec~\cite{NGAT4Rec} on Gowalla since they are not evaluated on Gowalla but only evaluated on the other two datasets in their original papers too, and the authors have not released their code. As the experimental settings of SGL-ED and NGAT4Rec are exactly same as us, we just report their results on Amazon-Books and Yelp2018. 
    
2) The performance of MF-CCL is surprising. When using CCL as the loss function, the performance of MF is not only much better than the results of MF-BPR reported in the previous paper, but also reaches a new state-of-the-art performance (if leaving out our SimpleX) on Amazon-Books and Yelp2018. On Gowalla, it also achieves comparable performance compared to the previous best model DGCF. Such results strongly suggest that loss functions can make a big difference and should be carefully chosen and studied.
    
3) YouTubeNet, CML, and SLIM are three models that we added and have not been tested on these three datasets before by the existing work. We found that they achieve pretty good performance. Specifically, these three models can averagely outperform a representative GNN-based CF model -- NGCF, by more than 24\% and 28\% w.r.t. Recall@20 and NDCG@20, respectively. This implies that the current baselines are relatively weak, which may lead us to overestimate how much real progress we have made in CF.
    
4) In CF tasks, more complex models not always lead to better performance. The designs of SLIM, YouTubeNet, CML, MF-CCL, and our SimpleX are all much more concise than most of autoencoder-based (e.g., Mult-VAE and MacridVAE) and GNN-based models (e.g., NGCF, NIA-GCN, and DGCF), but they can achieves better performance. This also reveals that the current trend in CF research, which pays too much attention to the design of sophisticated interaction encoders while ignoring the impacts of loss functions and negative sampling, needs to be improved.

\subsection{Incorporating CCL to Other Models}
In Table~\ref{fullresults1}, we have shown that one of the simplest models, i.e., MF, can even largely outperforms most of state-of-the-art models if training with CCL. We are curious about how other models will perform if incorporated with CCL instead of their original losses. 
Therefore, in this part, we take experiments with two effective CF models in addition, i.e., YouTubeNet and LightGCN with CCL, and report the results on Amazon-Books and Yelp2018 in Table~\ref{LightLossCompare}.
% We have the following findings from the results:
% \begin{itemize}
%     \item LightGCN-CCL performs very much better than original LightGCN-BPR on both Amazon-Books and Yelp2018. YouTubeNet achieves higher performance when training with CCL instead of SCE on all the three datasets. Such results further indicate that CCL is likely to be a more promising loss function to help CF models achieve better performance.
    
%     \item CCL does not make the performance of LightGCN better. On the contrary, the performance drops 5.8\% w.r.t. Recall@20.  
%  \end{itemize}

From the results, we find that training YouTubeNet and LightGCN with CCL instead of their original loss functions, i.e, SCE and BPR respectively, can bring good improvements. This demonstrates that CCL is likely to be a more promising loss function to help CF models achieve better performance.
Besides, we observe that the improvements brought by CCL on YouTubeNet and LightGCN are not as significant as those on MF. CCL seems to improve these models to a similar level of performance. 
This may be because of the following reason:
Generally, valuable collaborative information can be captured by both the interaction encoder and the loss function. As the encoders of YouTubeNet and LightGCN are sophisticated and stronger to learn biased collaborative signals, by contrast, the impact of the loss function to them appears relatively small.

In addition, it is worth noting that our main focus is to question the value of sophisticated encoders and provide a simple strong baseline, but not to improve current state-of-the-art CF models by exhaustingly trying of various loss functions.
Based on the experiments with MF, YouTubeNet, and LightGCN, we demonstrate and highlight that the loss function is a large bottleneck in CF models. We expect our work could inspire more research to study the co-design of the interaction encoder, loss function, and negative sampling.

\begin{table}[!t]
\caption{Performance of different models trained with CCL v.s. their original losses.}
\centering
\setlength{\tabcolsep}{0.48ex}
\begin{tabular}{c|cc|cc}
\hline
\multirow{2}{*}{Model} & \multicolumn{2}{c|}{Amazon-Books}  & \multicolumn{2}{c}{Yelp2018}  \\ \cline{2-5} 
                       & Recall@20       & NDCG@20         & Recall@20       & NDCG@20               \\ \hline
MF-BPR                 & 0.0338          & 0.0261          & 0.0549          & 0.0445              \\
MF-CCL                 & \textbf{0.0559} & \textbf{0.0447} & \textbf{0.0698} & \textbf{0.0572}  \\ \hline
YouTubeNet         & 0.0502          & 0.0388          & 0.0655         & 0.0537          \\
\hspace{-0.5ex}YouTubeNet-CCL         & \textbf{0.0544} & \textbf{0.0430} & \textbf{0.0685}               &  \textbf{0.0563}                         \\ \hline
LightGCN           & 0.0411          & 0.0315          & 0.0649          & 0.0530          \\
LightGCN-CCL           & \textbf{0.0528} & \textbf{0.0416} & \textbf{0.0669} & \textbf{0.0554}    \\ \hline
\end{tabular}
\label{LightLossCompare}
\end{table}

\subsection{Parameter Analysis on SimpleX}
\label{sec:ablation}
\begin{table}[!t]
\centering
\setlength{\tabcolsep}{1.1ex}
\caption{Parameter analysis results on SimpleX.}
\begin{tabular}{c|cc|cc}
\hline
\multirow{2}{*}{Ablations} & \multicolumn{2}{c|}{Amazon-Books}  & \multicolumn{2}{c}{Yelp2018}        \\ \cline{2-5} 
                 & Recall@20 & \hspace{-1ex}NDCG@20 & Recall@20 & \hspace{-1ex}NDCG@20 \\ \hline
avg\_pooling & \textbf{0.0583}
    & \textbf{0.0468}
  & \textbf{0.0701}
    & 0.0575
  \\
self\_attn. & 0.0580 
    & 0.0462 
  & 0.0698
   & \textbf{0.0576}
  \\
user\_attn. & 0.0551 
    & 0.0436 
  & 0.0698 
   & 0.0574   \\
 \hline
\hspace{-1.5ex}g = 0 & 0.0534    & 0.0429 &  0.0679   &  0.0555\\
g = 0.5 & \textbf{0.0583}
    & \textbf{0.0468}  & 0.0688  &  0.0565\\
\hspace{-1.5ex}g = 1 & 0.0540   & 0.0432 & \textbf{0.0701}
   & \textbf{0.0575}   \\
 \hline
$w$ = 1\quad\quad &  0.0163   &  0.0128 & 0.0238 
   &  0.0189 
\\
$w$ = 150\quad & 0.0542 

    & 0.0429 
 & \textbf{0.0701} 
  &  \textbf{0.0575}
\\
$w$ = 300\quad & \textbf{0.0583}
  & \textbf{0.0468} 
 & 0.0666 

   & 0.0549 
   \\
\hspace{1ex}$w$ = 1000 & 0.0481 
  & 0.0379 
 & 0.0568 

   & 0.0463 
  \\
 \hline
\end{tabular}
\label{ablationres}
\end{table}

We investigate the performance of three different behavior aggregation layers, the fusing weight $g$, and the negative loss weight $w$. Results on Amazon-Books and Yelp2018 are shown in Table~\ref{ablationres}. We can make the following observations: 1) Average pooling, self-attention, and user-attention obtain very similar results on Amazon-Books and Yelp2018, respectively. This shows the robustness of apply average pooling for behavior aggregation in practice. SimpleX with $g=0.5$ reaches higher performance compared with the other two settings on Amazon-Books, which shows that importance of fusing user embedding with user behavior aggregation. 2) The negative weight $w$ which adjusts the ratio of positive and negative losses is vital to model's performance. In general, too small ($w=1$) or too large ($w=1000$) difference between positive and negative losses leads to performance reduction. 
% The whole results of ablation study also indicate that the design choice of SimpleX should be data-driven. 

\begin{table}[!t]
    \centering
    \setlength{\tabcolsep}{1ex}
    \caption{Efficiency comparison on Amazon-Books, w.r.t. the average training time per epoch, the number of epochs to converge, and the total training time.}
\begin{tabular}{c|c|c|c}
\hline
Model       & Time/Epoch & \#Epochs & Training Time \\ \hline
\hline
ENMF        &      129s     &   81   &   2h54m        \\ 
LightGCN    &     51s   &   780   &   11h3m      \\ 
SimpleX ($|\mathcal{N}|$=100)    &     \textbf{40s}       &   \textbf{28}   &    \textbf{19m} \\
SimpleX ($|\mathcal{N}|$=1000)    &     131s       &   {35}   &    {1h16m}     \\ \hline
\end{tabular}
    
    \label{efficiencycompare}
\end{table}

%%%%%%%%%%%%%%%%%%%%%%%%%%%%%%%%%%%%%%%%%%%%%%
\begin{table*}[htbp]
\renewcommand\arraystretch{1.1}
\centering
%\small
% including NIA-GCN\cite{NIA-GCN}, BGCF~\cite{BGCF}, NBPO~\cite{NBPO}, DHCF~\cite{DHCF}, and LCFN~\cite{LCFN}. For fairness of comparison, we experiment with the same preprocessed data, and follow the embedding dimension and evaluation metrics as in the original work. The results of all the compared models are directly copied from the above five papers respectively.}
\caption{Performance comparison to some other models published in 2020. We use the same preprocessed datasets and report the same evaluation metrics with these papers. We also reuse their reported results for consistency.}
\begin{tabular}{c|cc||c|cc||c|cc}
\hline
\multicolumn{3}{c||}{Amazon-CDs} & \multicolumn{3}{c||}{Amazon-Movies}    & \multicolumn{3}{c}{Amazon-Beauty}              \\ \hline
Model        & Recall@20  & \hspace{-1.5ex}NDCG@20         & Model        & Recall@20 & \hspace{-1ex}NDCG@20 & Model        & Recall@20  & \hspace{-1ex}NDCG@20\hspace{-0.5ex} \\ \hline
NGCF      & 0.1258  & 0.0792         & NGCF        & 0.0866 & 0.0555    & MF-BPR       & 0.1312 & 0.0778 \\
NIA-GCN        & 0.1487 & 0.0932       & NIA-GCN        & 0.1058  & 0.0683   & NGCF       & 0.1513  & 0.0917  \\ 
BGCF        & 0.1506  & 0.0948          & BGCF       & 0.1066 & 0.0693   & BGCF      & 0.1534 & 0.0912 \\ \hline 
SimpleX  & \textbf{0.1763}  & \textbf{0.1145}  & SimpleX & \textbf{0.1342} &  \textbf{0.0926}  & SimpleX & \textbf{0.1721}  & \textbf{0.1028}  \\ \hline
\hspace{-1ex}RI over NIA-GCN & 18.6\%  & 22.9\%           & RI over NIA-GCN & 26.8\%       & 35.5\%    & --  & --  & --  \\
RI over BGCF & 17.1\%  & 20.8\%           & RI over BGCF & 25.9\%       & 33.6\%    & RI over BGCF & 12.2\%  & 12.8\%   \\\hline
\multicolumn{9}{c}{\vspace{-0.6ex}}\\

\hline
\multicolumn{3}{c||}{Amazon-Electronics} & \multicolumn{3}{c||}{CiteUlike-A}               & \multicolumn{3}{c}{Movielens-1M}              \\ \hline
Model        & F1@20  & \hspace{-1.5ex}NDCG@20         & Model        & \hspace{-0.5ex}Precision@20 & \hspace{-1ex}Recall@20 & Model        & F1@20  & \hspace{-1ex}NDCG@20\hspace{-0.5ex} \\ \hline
MF-BPR    & 0.0275 & 0.0680          & NGCF        & 0.0517       & 0.0193    & NGCF       & 0.1582 & 0.2511  \\
NBPO        & 0.0313 & 0.0810          & DHCF        & 0.0635       & 0.0249    & LCFN       & 0.1625 & 0.2603  \\ \hline 
SimpleX  & \textbf{0.0338} & \textbf{0.0842} & SimpleX & \textbf{0.0754} & \textbf{0.0269} & SimpleX & \textbf{0.1658} & \textbf{0.2670} \\ \hline
RI over NBPO & 8.0\%  & 4.0\%           & RI over DHCF & 18.7\%       & 8.2\%    & RI over LCFN & 2.0\%  & 2.6\%   \\ \hline
\end{tabular}
\label{fullresults2}
\end{table*}
%%%%%%%%%%%%%%%%%%%%%%%%%%%%%%%%%%%%%%%%%%%%%%%%%%%%%%

%%%%%%%%%%%%%%%%%%%%%%%%%%%%%%%%%%%%%%%%%%%%%%%%%%%%%%
\begin{table*}[htbp]
\centering
\renewcommand\arraystretch{1.1}
\setlength{\tabcolsep}{0.55ex}
\setlength{\belowcaptionskip}{-1ex}
%\small
% For fairness of comparison, we use the same preprocessed data and evaluation metrics as in the original work. We reduce the embedding dimension of items to 64 (i.e., d=64) for fair comparisons.
% \#Params shows the number of parameters used in each model.
\caption{Performence comparison to autoenoder-based models on Movielens-20M and MillionSongData.}
\begin{tabular}{c|cccc|cccc}
\hline
\multirow{2}{*}{Model} & \multicolumn{4}{c|}{Movielens-20M}   & \multicolumn{4}{c}{MillionSongData}                                                                                                      \\ \cline{2-9}
& \multicolumn{1}{l}{Recall@20} & \multicolumn{1}{l}{Recall@50} & \multicolumn{1}{l|}{NDCG@100} & \#Params 
& \multicolumn{1}{l}{Recall@20} & \multicolumn{1}{l}{Recall@50} & \multicolumn{1}{l|}{NDCG@100} & \#Params
\\ \hline
\multicolumn{1}{c|}{SLIM}                            & 0.370                         & 0.495                         & \multicolumn{1}{c|}{0.401}    & --   
      & --                         & --                         & \multicolumn{1}{c|}{--}     & --   
\\
\multicolumn{1}{c|}{Mult-VAE}                        & 0.395                         & 0.537                         & \multicolumn{1}{c|}{0.426}    & 24.5M 
      & 0.266                         & 0.363                        & \multicolumn{1}{c|}{0.313}    & 49.7M  
\\
\multicolumn{1}{c|}{EASE\textsuperscript{R}}         & 0.391                         & 0.521                         & \multicolumn{1}{c|}{0.420}    & 404.3M   
      & 0.333                        & 0.428                      & \multicolumn{1}{c|}{0.389}    & 1,692M   
\\
\multicolumn{1}{c|}{RecVAE}         & 0.414                         & 0.553                         & \multicolumn{1}{c|}{0.442}    & 16.5 M   
      & 0.276                        & 0.374                         & \multicolumn{1}{c|}{0.326}    & 33.3M   
\\ \hline
\multicolumn{1}{c|}{Mult-VAE (d=64)}                 & 0.375                         & 0.514                        & \multicolumn{1}{c|}{0.407}    & 2.6M  
      & 0.230                        & 0.319                       & \multicolumn{1}{c|}{0.280}    & 5.3M  
\\
\multicolumn{1}{c|}{EASE\textsuperscript{R} (d=64)}  & 0.361                         & 0.487                         & \multicolumn{1}{c|}{0.392}    & 2.6M   
 & 0.170                         & 0.235                         & \multicolumn{1}{c|}{0.205}    & 5.3M  
\\
\multicolumn{1}{c|}{RecVAE (d=64)}  & 0.385                         & 0.520                         & \multicolumn{1}{c|}{0.412}    & 2.6M   
 & 0.232                         & 0.319                         & \multicolumn{1}{c|}{0.280}    & 5.3M  
\\\hline
\multicolumn{1}{c|}{SimpleX (d=64)}                  & \textbf{0.389}                        & \textbf{0.523}                         & \multicolumn{1}{c|}{\textbf{0.416}}    & 1.3M 
  & \textbf{0.245}    & \textbf{0.329}                         & \multicolumn{1}{c|}{\textbf{0.293}}    & 2.6M 
    
    \\ \hline
\multicolumn{1}{l|}{RI over Mult-VAE}                & 3.8\%                         & 1.7\%                         & \multicolumn{1}{c|}{2.3\%}    &   
      & 6.5\%                         & 3.2\%                         & \multicolumn{1}{c|}{4.7\%}    &  
\\
\multicolumn{1}{l|}{RI over EASE\textsuperscript{R}} & 7.8\%                         & 7.4\%                         & \multicolumn{1}{c|}{6.2\%}    &   
 & 44.0\%                         & 40.3\%                         & \multicolumn{1}{c|}{43.3\%}    &          
\\ 
\multicolumn{1}{l|}{RI over RecVAE} & 1.1\%                         & 0.6\%                         & \multicolumn{1}{c|}{1.1\%}    &   
 & 5.8\%                         & 3.2\%                         & \multicolumn{1}{c|}{4.8\%}    &          
\\ 
\hline
\end{tabular}
\label{fullresults3}
\end{table*}
%%%%%%%%%%%%%%%%%%%%%%%%%%%%%%%%%%%%%%%%%%%%%%%%%%%%%%

\subsection{Efficiency Comparison}
Our SimpleX has high efficiency due to its simple design. We numerically compare the training time of SimpleX with two state-of-the-art CF models, i.e., ENMF and LightGCN, which are relatively efficient in their respective categories, on Amazon-Books. 
The efficiency experiments are conducted on the same Intel(R) Xeon(R) Silver 4210 CPU @2.20GHz machine with one GeForce RTX 2080 GPU.
We compare them under the same implementation framework, using the same acceleration methods (e.g., implementing the sampling with C++) to ensure fairness.
Specifically, we present the averaged training time per epoch, the number of epochs that the model needs to reach the level of performance reported in the original paper, and the total training time (test time is not included), in Table~\ref{efficiencycompare}.

% For ENMF, we use their official public source code~\footnote{\url{https://github.com/chenchongthu/ENMF}} for fair comparison.
% For LightGCN, we compare with its two official versions, i.e.,LightGCN (Pytorch)~\footnote{\url{https://github.com/gusye1234/LightGCN-PyTorch}} and LightGCN (Tensorflow)~\footnote{\url{https://github.com/kuandeng/LightGCN}}. The former uses multiprocess parallel sampling in Python and the latter adopts C++ to accelerate the negative sampling.

It turns out that SimpleX is much more efficient than ENMF and LightGCN overall.
Specifically, SimpleX only needs around 30 epochs to converge in training, which is more convenient for real application.
The total training time of SimpleX with a 1000:1 negative sampling ratio has around 2x and 10x speedup compared with ENMF and LightGCN respectively. Moreover, if we decrease the negative sampling ratio to 100:1, the training time for one epoch of SimpleX can be optimized to 40s, finally resulting in only 19 minutes total training time. Certainly, the performance slightly drops compared with using a 1000:1 negative sampling ratio, but it still maintains a pretty good level (much better than ENMF and LightGCN). Such high efficiency makes our model promising to be applied in large-scale real recommender systems.
% Note that if we apply C++ or parallelization for negative sampling as LightGCN, the efficiency can be further improved. Such high efficiency makes our model promising to be applied in large-scale real recommender systems.

% Furthermore, different from ENMF, SimpleX are final to obtain the embeddings of users and items, so they are suitable for fast KNN searching in the inference, supporting retrieving from millions of items in just milliseconds. Such higher efficiency makes our model promising to be applied in large-scale real recommender systems.

\subsection{Evaluating SimpleX on More Datasets}
In addition to the three main datasets used in the above sub-sections, we additionally evaluate SimpleX on 8 more datasets to further demonstrate the generability of SimpleX.

Table~\ref{fullresults2} shows the comparison results to some state-of-the-art CF models published in 2020. For fairness of comparison, we use the same data preprocessing and experimental settings (embedding dimensions and evaluation metrics) provided by the corresponding papers. We observed that SimpleX consistently outperforms all the compared models on different datasets. The performance improvements are especially large (12.8\% to 33.6\% improvement in NDCG@20) on Amazon-CDs, Amazon-Movies and Amazon-Beauty compared to BGCF, a recent GNN-based model. This again strongly verifies the effectiveness and robustness of SimpleX to serve as a strong baseline in future work.

Moreover, we also make a comparison to some autoencoder-based models, including SLIM, Mult-VAE, EASE\textsuperscript{R}, and RecVAE. It is worth noting that our experiment also follows the same setting with them. In particular, we adopt the strong generalization protocol, where the training, validation and test sets are disjoint in terms of users. This requires the model to perform inductive learning during inference. That is, only item embeddings can be learned during training and then transferred to the validation and test sets for prediction. To achieve this, we simplify SimpleX by setting $g=0$ in this experiment and only learn user representations from their historically interacted items.

%However, a large number of CF models, such as ENMF, CML, and most of GNN-based methods, as well as our SimpleX, are actually designed for transductive setting (i.e., weak generalization) since they have to learn the user embeddings. 
%Transductive setting is also more in line with the actual requirement of industry.

%Thus, to finish the comparison, we average the embeddings of the history interacted items of the user to represent the user embedding. 

Table~\ref{fullresults3} presents the evaluation results on Movielens-20M and MillionSongData. We can see that SimpleX obtains better performance than SLIM, which is a well-known strong baseline for CF. But it does not surpass Mult-VAE, EASE\textsuperscript{R} and RecVAE given their complete forms. This is reasonable because all of them use many more parameters ($\mathcal{O}(|I| * 600)$ for Mult-VAE and RecVAE, $\mathcal{O}(|I| * |I|)$ for EASE\textsuperscript{R}) than SimpleX, as shown in the ``\#Params" columns. Note that both Mult-VAE and RecVAE use 600 as the dimension of the first hidden layer. As the number of items ($|I|$) easily reaches  millions to billions in industrial recommender systems, we choose a small embedding dimension (i.e., 64) and results in parameters in the scale of $\mathcal{O}(|I| * 64)$. To make the comparison more fair, we reduce the embedding dimensions of baseline models accordingly. Specifically, for Mult-VAE and RecVAE, we set its encoder and decoder as a single $(|I| * 64)$-dimensional dense layer. For EASE\textsuperscript{R}, we decompose its item similarity matrix (denoted as B) to two $(|I| * 64)$-dimensional sub-matrices by truncated SVD, and multiply the two sub-matrices to approximate the item similarity matrix to perform predictions. In this setting, SimpleX clearly outperform these autoencoder based CF models.

Overall, our comprehensive experimental results on various datasets show that our SimpleX is simple and strong to serve as a new baseline model to facilitate future research on CF. The availability of this baseline would allow for more solid experimental evaluations and more fair comparisons among CF models.

%% file: sections/6_conclusion.tex
\section{Conclusion}\label{sec:conclusion}
In this paper, we study the progress made in CF research and identify three key aspects for CF modeling. While most research focuses on interaction encoders, the impacts of loss functions and negative sampling on CF models have been largely neglected. In this work, we highlight their impacts and further propose the cosine contrastive loss together with a simple and strong baseline for CF, dubbed SimpleX. It outperforms most state-of-the-art CF models by a large margin. Our work released the simple and strong baseline model and the whole benchmarking results for foster future research on CF. We conduct extensive experiments to validate the effectiveness and efficiency of SimpleX. We suggest that the CF community should pay more attention to other key components in addition to interaction encoders and encourage researchers to conduct more robust empirical evaluation.

\section{Acknowledgements}
This work was supported in part by the National Natural Science Foundation of China (61972219), the Research and Development Program of Shenzhen (JCYJ20190813174403598, SGDX20190918101201-696), the National Key Research and Development Program of China (2018YFB1800601), and the Overseas Research Cooperation Fund of Tsinghua Shenzhen International Graduate School
(HW2021013).